\documentstyle[multicol,epsf,aps,prb]{revtex}

\newcommand\ltdash{\raise-1.8pt\hbox{$\scriptscriptstyle |$}}
\newcommand \beq  {\begin{equation}}
\newcommand \eeq  {\end{equation}}
\newcommand \bea {\begin{eqnarray} }
\newcommand \eea {\end{eqnarray}}

\begin{document}
\draft
\twocolumn[\hsize\textwidth\columnwidth\hsize\csname @twocolumnfalse\endcsname
\title{
Local Moments in an Interacting Environment
}
\author{ P. Coleman 
$^{*}$
and A. M. Tsvelik}
\address{Department of Physics,
Oxford University
1 Keble Road
Oxford OX2 3NP, UK
}
\date{\today}
\maketitle
\begin{abstract}

We discuss how local moment physics is modified
by the presence of interactions in the
conduction sea.
Interactions in the conduction sea are shown to
open up new symmetry channels for the
exchange of spin with localized moments.
We illustrate this
conclusion in the strong-coupling limit by carrying out a Schrieffer
Wolff transformation for a local moment in an interacting electron
sea, and show that these corrections become very severe in the
approach to a Mott transition.
\end{abstract}
\vskip 0.2 truein
\pacs{78.20.Ls, 47.25.Gz, 76.50+b, 72.15.Gd}
\newpage
\vskip2pc]

\section{Introduction}

In recent times, the ``quantum chemistry'' approach
has proven one of the most effective ways to  formulate
minimal models of strongly correlated electron systems.
The corresponding
strategy of first solving the physics of a strongly interacting
atom or cluster,  and later 
superimposing the  inter-site couplings
has provided an  underlying philosophy
for many  models of interacting electron systems,\cite{anderson,hubbard}
and led to several new concepts, such as the ``local
moment''\cite{anderson}, the ``upper'' and ``lower'' Hubbard
bands\cite{hubbard} and the ``Zhang Rice singlet''.\cite{zhang}

A key underlying assumption of the quantum chemistry
approach is  that 
the interacting environment which develops
around each  local scattering center, atom
or cluster, does not qualitatively change its scattering properties.
This long-held assumption 
may not hold in all 
densely interacting systems and for this reason,
deserves
special scrutiny.
We already know that this assumption fails in one
dimension, where 
interactions in the bulk Luttinger liquid alter the scaling exponents
for forward and backward
scattering,  qualitatively
changing the character of the scattering center. A weak potential
scatterer renormalizes into an infinitely strong blockade to
transport,\cite{fisher} whilst a one-channel Kondo develops properties
reminiscent of a two-channel Kondo effect.\cite{dhlee,furuk}

Motivated by these considerations,
this paper discusses how an interacting environment can 
qualitatively modify the scattering
properties of a local moment in higher dimensions.  
In one dimension,  forward and backward scattering are delineated
by their effects on spin-charge coupling: the former
preserves spin-charge decoupling, whereas the latter
couples spin and charge together.  This accounts
for their very different scaling properties in the presence of
interactions.  In higher dimensions, spin exchange between a local
moment and its environment 
can be similarly divided, and in keeping with the lower
dimensional analog, Coulomb interactions tend to suppress those
components of the spin scattering that couple to charge currents.
Some aspects of these effects have been discussed by Schork and
Fulde\cite{schork}.  Our paper serves to highlight a  particular
point, namely
that this effect gives rise to
new spin-exchange channels between the local moment and its
environment. 
In the lattice, these 
new scattering channels qualitatively modify the interactions 
between mobile Kondo singlets. A forthcoming paper\cite{tobe}
will discuss how second channel scattering in a Kondo lattice
can give rise to a collective Kondo effect that destabilizes the 
Fermi liquid and ultimately gives rise to composite
pairing.\cite{zachar}

\section{Magnetic Impurity in a non-interacting Environment}

The usual starting point for studying a magnetic impurity is
the Anderson impurity model\cite{anderson}.    We  shall examine
how the reduction of the Anderson model to a Kondo model is affected
by the presence of interactions amongst the conduction electrons. 
We begin with a brief resum\'e of the situation in a non-interacting
environment.
The original Anderson model is written
\bea
H = H_o + H_{v} + H_{d}
\eea
where
\bea
H_o = \sum_{{\bf k}\sigma} \epsilon_{\bf k}  c{^{\dag}}_{{\bf k}\sigma}c_{{\bf
k}\sigma}
\eea
describes a sea of conduction electrons, 
\bea
H_d = E_d d{^{\dag}}_{\sigma}d_{\sigma} +  U n_{d \uparrow}n_{d\downarrow},
\qquad(n_{d\sigma} = d{^{\dag}}_{\sigma} d_{\sigma}),
\eea
is the Hamiltonian for a localized d-state, with an on-site Coulomb
interaction of strength $U$, and 
\bea
H_{v} = 
V\sum_{{\bf k} \sigma}
[\Phi_{d{\bf k}}c{^{\dag}}_{{\bf k}\sigma}d_{\sigma}  + {\rm H. c.} ].
\eea
describes the hybridization between the continuum and the localized
atomic orbital. 
The matrix element
\bea
V\phi_{d{\bf k}} = \int d{\bf x} e^{i \bf k \cdot x}V({\bf x})
\phi_d({\bf x})
\eea
is the overlap of the local orbital with the surrounding
conduction electron orbitals. 
An important point to note is that the local atomic orbital
only hybridizes with a {\em single} Wannier state with a particular local
symmetry.  For a transition metal system,
$\phi_{d{\bf k}}$ has d- symmetry, in a heavy
fermion system this matrix element has f-symmetry.
The single-channel nature of the model becomes clear in
a tight-binding
representation, for if
$c{^{\dag}}_{j \sigma} = \sum_{\bf k} c{^{\dag}}_{\bf k \sigma}e^{-i{\bf k} \cdot {\bf
x_j}}$ creates an electron at site j, then
$
\Phi_{d \bf k}= \sum_j \Phi_d({\bf x_j})e^{-i {\bf k}\cdot {\bf x_j}}
$
is clearly
the form-factor of a  Wannier state of nearby atomic orbitals so that
$
\psi^{\dag}_{d \sigma} =  \sum_j \Phi_d({\bf x_j})c^{\dag}_{j\sigma}
$
creates an electron at this state.
In this basis
the hybridization 
can be written
\bea
H_{v} = 
V\sum_{ \sigma}[d{^{\dag}}_{ \sigma} \psi_{d\sigma} + {\rm H. c.} ].
\eea

A large Coulomb interaction $U$ suppresses charge fluctuations
on the impurity site, causing local moment formation in the
``d-orbital''.  \cite{anderson}
In this situation, virtual charge fluctuations
induce 
an anti-ferromagnetic interaction between the local moment
and the surrounding conduction sea and the Anderson model can
be further reduced by means of 
a Schrieffer-Wolff\cite{schrieffer}
transformation which integrates out these fluctuations to yield
an effective Kondo model
\bea H = H_o + H_I
\eea
where
\bea
H_I = J {\bf S} \cdot \psi{^{\dag}} _{d}
\pmb{$\sigma$}\psi_{d}
\eea
describes the residual anti-ferromagnetic interaction between the
spin of the local moment ${\bf S} = \frac{1}{2}d{^{\dag}}{\pmb{$\sigma$}}d$
($n_d=1$) and the electron spin-density and
\bea
J =  \left( {V^2 \over U+E_d}
\right) +
\left({V^2 \over -E_d}
\right)\label{J}
\eea
where $E_d$ is taken to be negative.
The two terms in this expression are the 
perturbations to the energy resulting from 
virtual charge fluctuations $d^1+e^-\rightleftharpoons d^2$ and
$d^1\rightleftharpoons d^0+ e^-$
into the  $d^2$ and $d^0$
configurations respectively.  Once again,  the local moment only
interacts with a single Wannier orbital.

In momentum space the Kondo
interaction  can be written 
\bea
H_I= \sum_{{\bf k}, {\bf k'}}
J_{{\bf k}, \bf k'} c{^{\dag}}_{{\bf k}} {\pmb{$ \sigma$}}c_{{\bf k'}} \cdot {\bf S}
\label{ham}
\eea
where 
\bea
J_{{\bf k}, \bf k'} = J \Phi_{d{\bf k}}\Phi^*_{d{\bf k'}},
\label{ho}
\eea
involves a single Wannier state. 
In a site basis, the Kondo interaction becomes
\bea
H_I= \sum_{l,l'}J_{l,l'}
 c{^{\dag}}_{l} {\pmb{$ \sigma$}}c_{l'} \cdot {\bf
S}.\label{nonloc}
\eea
where 
$J_{l,l'} = J  \Phi_d({\bf x_l})\Phi^*_d({\bf x_{l'}})$.
The non-locality of the exchange means that an 
electron at a neighboring orbital can exchange spin with the local
moment at the same time as hopping to one of the other neighboring
orbitals.  These are the processes which couple spin and charge
fluctuations together. 

\section{Effect of interactions in the environment}

\begin{figure}[tb]
\epsfxsize=3.in 
\centerline{\epsfbox{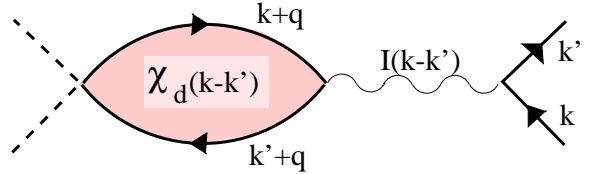}}
\vskip 0.4truein
\protect\caption{Vertex correction to Kondo interaction.}
\label{Fig1}
\end{figure}

Now let us discuss how the spin exchange between the local 
moment and its environment is modified when the surrounding
environment becomes interacting.   Suppose we introduce a weak
spin-spin interaction into the conduction sea , writing
\bea
H = H_o + H_I + \sum _{\bf q} I({\bf q}) \pmb{$\sigma$}_
{\bf -q}\cdot \pmb{$\sigma$}_{\bf q}
\eea
where $\bf \pmb{$\sigma$}_{\bf q} = \sum_{\bf k }
 c{^{\dag}}_{{\bf k- q}} \pmb{$ \sigma$} c_{{\bf k}}
$ is the conduction electron spin-density at momentum $\bf q$
and $I({\bf q})$ defines the strength of spin-spin interactions
at this wavevector. 
To leading order $O(I)$, there is 
a vertex correction to the Kondo interaction, as shown
in Fig \ref{Fig1}.  Written out explicitly, this gives
\bea
J_{\bf k, k'} = J^{(o)}_{\bf k, k'}  + J \chi_d({\bf k}-{\bf k}'
) I({\bf k}-{\bf k}')
\eea
where 
\bea
\chi_d({\bf q})= 2 \sum_{\bf k} \frac{f(\epsilon_{\bf k-q})
- f(\epsilon_{\bf k})}
{\epsilon_{\bf k}
- \epsilon_{\bf k-q}} \Phi^*_{d{\bf k-q}}\Phi_{d{\bf k}},
\eea
is the spin-susceptibility of the d-state to a 
magnetic field at wave-vector $\bf q$.
By expanding the  Kondo coupling in terms of a complete
set of orthogonal Wannier states $\{ \Phi_{\lambda{\bf k}}\}$
with crystal field symmetry $\lambda$, 
$
J_{\bf k, k'} = \sum_{\lambda}J_{\lambda}
\Phi_{\lambda{\bf k}}\Phi^*_{\lambda{\bf k}'}
$, we see that 
\bea
J_{\lambda}= J\delta_{d\lambda} + J \sum_{{\bf k}, \bf k'}
\chi_d({\bf k}-{\bf k}') I({\bf k}-{\bf k}')\Phi^*_{\lambda{\bf k}}\Phi_{\lambda{\bf k'}}.
\eea
now contains components in new symmetry channels $\lambda \neq d$.

\begin{figure}[tb]
\epsfxsize=3.0in 
\centerline{\epsfbox{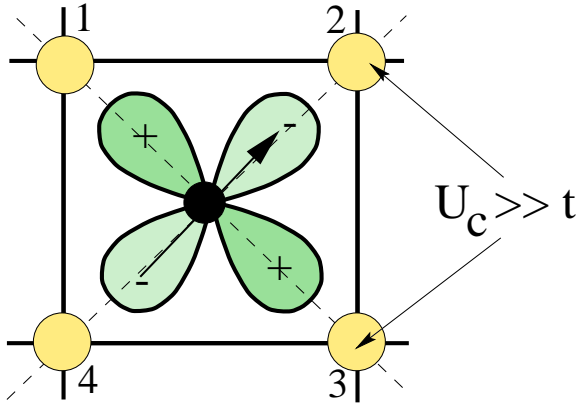}}
\vskip 0.4truein
\protect\caption{Magnetic moment in an interacting environment.
Localized electron at center of plaquet hybridizes in the $d_{xy}$-channel
with  nearby atoms.  The onsite interaction at each atomic
site $U_c$ is taken to be far larger than the electron band-width $t$.
}
\label{Fig2}
\end{figure}

To follow how these effects grow with the strength of interaction,
we now repeat the analysis in the strong coupling limit, 
carrying out a Schrieffer-Wolff transformation in the presence
of a strongly interacting environment. 
To be specific, 
consider
a two-dimensional  tight-binding
model of conduction electrons  
with a local
moment located in the center of a single  square plaquet at the
origin (Fig. \ref{Fig2}). 
If the onsite Coulomb interaction between the electrons
on the lattice is much larger than the band-width, 
the motion of the electrons is described by an infinite U
Hubbard model\cite{hubbard}
\bea
H_o = \sum_{l,l', \sigma} [ t_{l l'} - \mu \delta_{l l'} ] {\rm X}^{\dag}_{l \sigma} {\rm X}_{l'\sigma}
\eea
where ${\rm X}_{j\sigma} = c_{j \sigma} ( 1 - n_{j - \sigma})$
is a Hubbard operator \cite{hubbard} and $t_{l l'}= -t$ for nearest
neighbors, but is zero otherwise. 
Suppose that the localized state has a d-symmetry, 
so that
\bea 
H = H_o + 
H_v + H_d
\eea
where 
\bea
H_{v} = 
V\sum_{ l \sigma}[ \Phi_d(x_l)d^{\dag}_{ \sigma}{\rm X}_{l\sigma} 
+ {\rm H. c} ],
\eea
and
With the configuration shown in (Fig \ref{Fig2}),
it is the $d_{xy}$ orbital of the local moment which hybridizes
strongly with the nearby atoms. The value of the 
$d_{xy}$ Wannier state at the
four nearest-neighbor
sites labeled sequentially around the local
moment (Fig \ref{Fig2}) is
\bea
\Phi_{\lambda} [{\bf R}_i]
 = \frac{1}{2}(1,\ -1,\ 1,-1),
\eea
where  $\Phi({\bf x})=0$ for more distant neighbors.
In the limit where $|E_d|$ and $U+E_d$ are much larger
than both $t$ and $V$, only virtual charge fluctuations take place at the 
localized moment. We may integrate these fluctuations out by
carrying out a Schrieffer Wolff transformation
$H\longrightarrow H^* = e^{i S} H e^{-iS}$ where
$S$ is chosen to eliminate the hybridization term, 
$i[S,H_o]= -H_v$. This yields
\bea
H ^* = H_o + H_I 
\eea
where  
\bea
H_I=
\left\{
\frac{V^2}{E_d}AA^{\dag} 
-\frac{V^2}{U+E_d}A^{\dag} A
\right\},
\eea
where 
\bea
A = \sum_l \Phi_d(x_l)d^{\dag}_{\sigma}{\rm X}_{l\sigma }.
\eea
Re-ordering the operators, we find that
\bea
H_I = 
J( {\bf S} 
\cdot \Psi{^{\dag}}_{d } \pmb{$\sigma$}
\Psi_{d})-K
(\Psi{^{\dag}}_{d } \Psi_{d}
)
\eea
where $J$ is given by (\ref{J}),
\bea
\Psi_{d \sigma }= \sum_l \Phi_d({\bf x_l}){\rm X}_{l \sigma},
\eea
and
\bea
K =  \left( {V^2 \over U+E_d}
\right) +\left({V^2 \over E_d}
\right).
\eea
For simplicity, 
we chose the symmetric case, where $U+E_d = -E_d$ so $K=0$
and potential scattering vanishes. In this case 
the interaction between the local moment and its environment 
takes the form 
\bea
H_I& =&  \sum_{l,l'}J_{l,l'}{\bf S} 
\cdot {\rm X}{^{\dag}}_{l } \pmb{$\sigma$}
{\rm X}_{l'},
\eea
where $J_{l,l'} = J/4$ for all sites around the spin.  We see that the
net effect of the strong interactions in the environment is
to replace the conduction electron operators by Hubbard operators
\bea
c_{j\sigma} \longrightarrow c_{j \sigma}(1 - n_{j -\sigma}) = {\rm X}_{j \sigma}, 
\eea
We now examine the consequences of this replacement. 

We may divide the Kondo interaction into a one-site and two-site
component, writing
\bea J_{l,l'} = (J/4)[\delta_{ll'} + (1-\delta_{l,l'})].
\eea
These two terms are the {\em loose} analog of forward, and backward
scattering in one dimension.  The site diagonal 
terms
do not involve charge fluctuations and these are unaffected
by the presence of interactions. 
By contrast, 
processes where
the electron exchanges spin and hops from site to site
are suppressed by the Coulomb interactions in the conduction sea:
these processes are completely eliminated
in the limit where there is one electron per site.

We may
make a crude estimation of the effect of the Hubbard operators
by making a Gutzwiller approximation:
\bea
{\rm X}{^{\dag}}_{j } \pmb{$\sigma$}
{\rm X}_{l}\longrightarrow 
c{^{\dag}}_{j} \pmb{$\sigma$}
c_{l}
\times \left\{
\begin{array}{cr}
1,&(j=l)\cr
(1-x)
,&(j \ne l)\end{array}\right.
\eea
where $x$ is the concentration of electrons.
This approximation yields the right physics for $x\sim 0$ and
in the limit $x\rightarrow 1$.
It follows that
\bea
H_I &=&\sum_{l, l'}
J_{ll'}
{\bf S} 
\cdot c{^{\dag}}_{l } \pmb{$\sigma$}
c_{l'},\cr
J_{l,l'}&=& \frac{J}{4}[(1-x)+x\delta_{ll'}]
\eea
The first term in $J_{l,l'}$ describes spin exchange in the
original single channel.  The second term is site-diagonal
and therefore involves a sum over new spin exchange channels.
For this lattice 
there are four orthogonal Wannier states 
$\Phi_{\lambda}$, $\lambda = (1,4)$ which overlap with the
nearest neigbor atoms.  The value of the Wannier
state at the four sites labeled sequentially around the local
moment is then
\bea
\Phi_{\lambda} ({\bf R}_i)
 = \frac{1}{2}(1,-i^{\lambda},(-1)^{\lambda}
, -(-i)^{\lambda})
\eea
where we identify $\Phi_0 \equiv \Phi_d $, with the
primary d-channel.  $\lambda =1$ and $\lambda = 3$ correspond to
p-channels, whereas $\lambda=3$ describes the extended s-channel.
If we expand $J_{l,l'}$ in this basis, writing
$J_{\lambda }= \sum_{l,l'}J_{l,l'}\Phi^*_{\lambda}({\bf x_l})
\Phi_{\lambda}({\bf x_{l'}})$, we find that
\bea
J_{\lambda}/J = 
\left\{ 
\begin{array}{crcll}
1 - \frac{3x}{4},& \lambda&=&0,\  & \hbox{Primary ch.}\cr
\cr
\frac{x}{4}, & \lambda &=&1,2,3, & \hbox{Secondary ch.}
\end{array}\right.
\eea
so that interactions induce spin exchange in three new channels:
two p-  and one extended s-channel, each 
with scattering amplitude $Jx/4$.  Schematically
\[
\hbox{d-channel}
\stackrel
{\rm interactions}
{\hbox to 30pt{\rightarrowfill}}
\hbox{ d, p, s-channel}
\]
We may compactly represent the
spin-exchange by replacing 
$J_{\bf k, \bf k'}$
 in (\ref{ho}) by
\bea
J_{\bf k, \bf k'} = \sum_{\lambda = 0,3}J_{\lambda}\Phi_{\lambda{\bf k}} \Phi^*_{\lambda{\bf k'}}
\eea
\enlargethispage{-3\baselineskip}
Remarkably, the strength of the scattering in the other channels
is broadly comparable with that in the primary channel, and
in the extreme limit of one electron per site ($x=1$),
the amplitude to scatter becomes equal in each channel. In this
special limit, all spin-hop processes have been suppressed, and
the Kondo interaction becomes four individual Heisenberg
spin couplings to each neighboring atom. 
This means that in 
the vicinity of a Mott transition,
a local moment will behave as a multi-channel Kondo model.

\section{Channel Symmetry and implications for the Kondo Lattice}

Physical realizations of a Kondo lattice will always involve
electron interactions in the conduction sea. From the
arguments we have just developed, we expect these interactions
to induce a Kondo coupling in new symmetry
channels. Predominantly f-channel heavy fermion systems
are expected to develop weaker spin-exchange couplings to 
the d, p and s-channels. Likewise, d-channel transition metal systems
will develop weaker Kondo coupling to the p and s-channels.

At first sight,
these weaker secondary  couplings might be thought to be irrelevant, as they
are for example, in a single impurity model\cite{blandin,jerez,coxrev}.
For an impurity magnetic ion, the Kondo effect develops exclusively in the
strongest screening channel.  
However, Kondo impurity models have
a special local symmetry which preserves the channel quantum number of
scattered electrons.  By contrast, an electron travelling 
in a Kondo lattice can change symmetry channels as it moves
from one spin site to another, so that 
channel quantum number is not conserved.  
This has a 
profound influence on the Kondo lattice, for it means that the
the subspace of Kondo singlets in one channel is no longer
orthogonal to the 
subspace of Kondo singlets in other channels. Thus 
the development
of Kondo effect in one channel no longer excludes the possibility
of a Kondo effect developing coherently in the other channels.

To illustrate this point we  shall consider
a two-channel Kondo lattice in the strong coupling limit,
where the band-width
is set to zero, so
\bea
H&=& H^{(1)}+H^{(2)}\cr
H^{(\lambda)}&=& (J_{\lambda}/N_s)\sum_{{\bf k}, {\bf k'},j}
\Phi_{ \bf \lambda k}
\Phi^*_{ \bf \lambda k'}
 c{^{\dag}}_{\lambda {\bf k}} {\pmb{$ \sigma$}}
c_{\lambda {\bf k'}} \cdot{\bf S}_j
e^{i ({\bf k' } - {\bf k}).{\bf R}_j},
\label{ham2}.
\eea
where $\sum_{\bf k} \Phi_{1 \bf k}\Phi^*_{2 \bf k} =0$
defines the orthogonality between the channels and $N_s$ is the
number of sites in the lattice. 

\begin{figure}[tb]
\epsfxsize=3.0in 
\centerline{\epsfbox{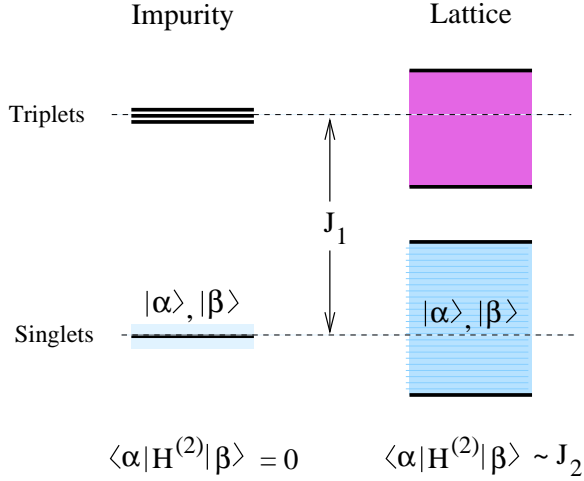}}
\vskip 0.1truein
\protect\caption{Contrasting the strong-coupling
limit of a single impurity and lattice model
with a weak second-channel coupling. In the impurity
model, there is no matrix element of $H^{(2)}$
in the low-energy subspace. In the lattice, where
channel number is not conserved, the matrix element
of $H^{(2)}$ in the low-energy subspace is finite,
and gives rise to interactions amongst the mobile
Zhang-Rice singlets.
}
\label{Fig3}
\end{figure}

Let us now
contrast the effect of $H^{(2)}$ in a single
impurity model, with its effect in a lattice. (Fig. \ref{Fig3})
Suppose $J_1 >> J_2 $, so that the low-energy physics
is determined by the projection
of $H$  into the space of
Kondo singlets in channel one. First consider
an impurity model.
For $J^{(2)}=0$, the
ground-state is a Kondo singlet formed between 
the local moment, and an electron in channel one
\bea
\vert \phi \rangle =
\frac{1}{2}[\psi^{\dag}_{1 \uparrow}d^{\dag}_{\downarrow]}-
\psi^{\dag}_{1 \downarrow}d^{\dag}_{\uparrow}]\vert 0\rangle,
\eea
where 
$\psi^{\dag}_{1\sigma} = N^{-1/2}\sum_{\bf k} \Phi_{1 \bf k} c{^{\dag}}_{{\bf k}\sigma}$, and 
we have represented
 ${\bf S} = d^{\dag}\bigl( \frac{\pmb{$\sigma$}}{2}\bigr)d$. 
Now 
$H^{(2)}$ flips
the spin of the local moment without affecting the  spin of
the electron it is bound to. To see this,  note that 
$
H^{(2) }
= J_2 \bf S \cdot 
\psi^{\dag}_2 \pmb{$\sigma$}\psi_2
$
where  $\psi^{\dag}_{2} = N^{-1/2}\sum_{\bf k} \Phi_{2 \bf k} c{^{\dag}}_{{\bf k}}$. Orthogonality of the scattering channels 
guarantees that $\{\psi_{1 \sigma},\
\psi^{\dag}_{2\sigma'} \} =0$, i.e $\psi_{2}$ has no overlap with 
the bound-electron in channel one. This means that
when we project $H^{(2)}$ into the low-energy 
subspace, 
\bea
H^{(2)} \rightarrow   \langle \phi \vert {\bf S} 
\vert \phi \rangle \cdot 
\psi^{\dag}_2
 \pmb{$\sigma$}\psi_2
=0
\eea
because there are no matrix elements of the spin operator
$\bf S$ in the singlet subspace.

By contrast, in the lattice where 
channel conservation is lost, $H^{(2)}$  
does act on the electrons bound into Kondo singlets, so that 
there are 
finite matrix elements of $H^{(2)}$ in the low-lying singlet
subspace of channel one.
If $\vert \alpha\rangle$ and $\vert \beta \rangle $ are states
in this low-lying subspace, this means
\bea
\langle\alpha \vert H^{(2)}\vert \beta \rangle = \left\{
\begin{array}{cc}
0,& \hbox{impurity}\cr
&\cr
O(J_2), & \hbox{lattice}
\end{array}
\right.
\eea
This 
marks a qualitative difference between the impurity
and lattice models. It means that we can no longer  tacitly assume that
in the lattice
second-channel couplings are an irrelevant perturbation.

We now calculate the form of these
additional terms in the lattice.  We follow
the method developed by Zhang and
Rice for reducing a two-band model of the cuprate perovskites
to a one-band t-J model\cite{zhang}. The Zhang-Rice reduction
to a single band was carried out on a model with spin-exchange
in a single ($d_{x^2-y^2}$) channel. We now examine how this
analysis changes when a weak
  additional  spin exchange channel is introduced.
We first construct a set of orthogonal
Zhang-Rice singlet operators for channel one.  
An electron in the Wannier state
with the symmetry 
of channel one is created at site $j$ by the operator
\bea
p^{\dag}_{j\sigma} = \frac{1}{\sqrt{N}}
\sum_{\bf k} \frac {\Phi_{1 \bf k}}{|\Phi_{1 \bf k}|}
 e^{i {\bf k \cdot R_j}}
c^{\dag}_{\bf k\sigma}
\eea
We can write both $H^{(1)}$ and $H^{(2)}$ in this basis
as follows
\bea
H^{(\lambda)} = \frac{J_{\lambda}}{N_s}\sum_{{\bf k}, {\bf k'},j}
\tilde \Phi_{\lambda \bf k}
\tilde \Phi^*_{ \lambda\bf k'}
p{^{\dag}}_{ {\bf k}} {\pmb{$ \sigma$}}
p_{ {\bf k'}} \cdot{\bf S}_j
e^{i ({\bf k' } - {\bf k}).{\bf R}_j}.
\eea
where 
\bea
\tilde \Phi_{\lambda \bf k } =
|\Phi_{\lambda \bf k }|
\biggl[
{\Phi^*_{ 1 \bf  k}
\Phi_{ \lambda \bf  k}\over | \Phi_{1 \bf k} || \Phi_{\lambda \bf k} | }
\biggr].
\eea
Our ability to write $H^{(2)}$ in terms of the Wannier states of channel
one is a direct consequence of the absence of channel conservation. 

The low-lying basis of Zhang-Rice singlets for $H^{(1)}$ is constructed
using the operator
\bea
b^{\dag }_j = 
\frac{1}{\sqrt{2}}
[p^{\dag}_{j \uparrow}d^{\dag}_{j\downarrow}-
p^{\dag}_{j \downarrow}d^{\dag}_{j\uparrow}],
\eea
to creat a ``Zhang-Rice'' singlet in channel one at site $j$. 
In the low-lying manifold of states, each site is either occupied by
a Zhang-Rice singlet, or an isolated d-spin.  The  vacuum  corresponds
to a singlet at every site
\bea
\vert \phi\rangle = \prod_j b^{\dag}_j \vert 0\rangle,
\eea
and a general state is formed by acting on this state with the Hubbard operator
${\rm X}^{\dag}_{j\sigma} = \sqrt{2} d^{\dag}_{j \sigma}b_j$ as follows
\bea
\vert \{j \sigma_j\}\rangle = 
\prod _{\{j, \sigma_j\}} {\rm X}^{\dag}_{j\sigma_j}
\vert \phi\rangle.
\eea
Within this manifold of states an electron can only be added
by the creation of a Zhang-Rice singlet. For states
$\vert \alpha \rangle, \vert \beta \rangle $ that lie
in the low-lying subspace
$\vert \{j \sigma_j\}\rangle $,  
\bea
\langle \alpha \vert
p^{\dag}_{j \sigma}
\vert \beta \rangle
=\langle \alpha \vert
\sqrt{2}\sigma b^{\dag}_j d_{j, - \sigma}
\vert \beta \rangle
=
\langle \alpha \vert
\sigma {\rm X}_{j - \sigma}
\vert \beta \rangle\eea
so we may carry out the projection into the low-energy subspace
by replacing $p^{\dag}_{j\sigma} \rightarrow \sigma {\rm X}_{j\ - \sigma}$.
The projected form for $H^{(\lambda)}$ is then
\bea
H^{(\lambda)} = \frac{J_{\lambda}}{N_s}\sum_{l,l',j}
\tilde \Phi_{\lambda}({\bf x}_{l'j})
\tilde \Phi^*_{\lambda}({\bf x}_{lj})
{\rm X}{^{\dag}}_{l'} {\pmb{$ \sigma$}}
{\rm X}_{l} \cdot{\bf S}_j.
\eea
On the sites where $l=j$ or $l'=j$, we can use the identity
$(
{\bf S}_j\cdot \pmb{$\sigma$})
 {\rm X}_j = - \frac{3}{2}{\rm X}_j$, to obtain
\bea
H^{(\lambda)} &=&
\sum_{\bf i,j} t^{\lambda}_{ij}
{\rm X}^{\dag}_{i \sigma } {\rm X}_{j \sigma} \cr
&+& \frac{J_{\lambda}}{N_s}\sum_{l, l' \ne j}
\tilde \Phi_{\lambda}({\bf x}_{l'j})
\tilde \Phi^*_{\lambda}({\bf x}_{lj})
{\rm X}{^{\dag}}_{l} {\pmb{$ \sigma$}}
{\rm X}_{l '} \cdot{\bf S}_j,
\label{ham3}
\eea
where $t^{\lambda}_{jl}=
N_s^{-1}\sum_{\bf k} t_{\lambda}({\bf k})
e^{i {\bf k} \cdot {\bf R_{jl}}}
$ and
\bea
t_{\lambda}({\bf k}) = -3  J_\lambda  
\tilde \Phi_{\lambda}(0) {\rm Re}[\tilde  \Phi_{\lambda \bf k}]
\eea
The first term in (\ref{ham3})
describes the motion  of the  Zhang-Rice holes. 
In general,  $\tilde \Phi_{2 \bf k}$ is a function with nodes,
so  $\tilde \Phi_{2}(0)$ vanishes, and 
$H^{(2)}$ contributes solely to an anisotropic interaction amongst
the holes. 
\bea
H^{(2) }= J_2 \sum_{l, l',j}
\tilde \Phi_{2}({\bf x}_{l'j})
\tilde \Phi^*_{2}({\bf x}_{lj})
{\rm X}{^{\dag}}_{l} {\pmb{$ \sigma$}}
{\rm X}_{l '} \cdot{\bf S}_j,
\eea
The symmetry of this term is governed by the product of form-factors
$\tilde \Phi_{2\bf k} \propto \Phi_{2\bf k} \Phi^*_{1\bf k}
$, 
a function that has to contain nodes, because of the orthogonality
of form-factors ($\sum _{\bf k}\Phi_{2\bf k} \Phi^*_{1\bf k} =0$ ).
In the primary channel, the corresponding interaction term
has an isotropic ``extended-s'' symmetry.
This term
is numerically small 
and is generally neglected as an irrelevent
perturbation to the
infinite (s-wave) onsite repulsion between holes.
The final form for the effective 
Hamiltonian is
\bea
H = t\sum_{(i,j)} 
{\rm X}^{\dag}_{i } {\rm X}_{j } 
+ \frac{J_{2}}{N_s}\sum_{j, {\bf a, \bf a'}}
\tilde  \Phi_{2}({\bf a})
\tilde  \Phi^*_{2}({\bf a'})
{\rm X}{^{\dag}}_{j+{\bf a}} {\pmb{$ \sigma$}}
{\rm X}_{j+{\bf a'}} \cdot{\bf S}_j,\nonumber
\eea
where, we have neglected all but the nearest neighbor
coefficients, so that $(i,j)$ represent nearest neigbors,
$\bf a$ is a vector linking nearest neigbors, 
$ t = - 3 \tilde \Phi_1(0)\tilde \Phi_1({\bf a})J_1$.
The second term shows that
 spin-exchange processes in channel two survive the projection into
the subspace of singlets for channel one. For this reason, we can no 
longer expect singlet formation in one channel to pre-empt 
a Kondo effect in the second, weaker channel.

One of the interesting possibilities that this presents
us with, is the possibility that Kondo spin-exchange in the second-channel
can generate pairing.  If
we consider a pair of Zhang-Rice
holes, then the matrix elements between
the two states produced by $H^{(2)}$ is given by
\bea
\langle {\bf k}\uparrow,-{\bf k}
\downarrow
\vert H^{(2)}\vert
{\bf k}'\uparrow,-{\bf k}'\downarrow
\rangle\propto -J_2 \tilde \Phi_{2\bf k}\tilde \Phi^*_{2\bf k'}.
\eea
In 
the original Zhang-Rice problem, the primary
spin-exchange channel has  $d_{x^2 - y^2}$ symmetry.
The projected form factor for the
primary $d_{x^2 - y^2}$ spin-exchange channel is\cite{zhang}
\bea
\tilde \Phi_{1 \bf k}= (1 + \frac{1}{2}[\cos(k_x) + \cos(k_y)])^{\frac{1}{2}}
\eea
We expect
there to also be spin-exchange terms of strength $J_2 \sim (\delta/8) J_1$,
where $\delta$ is the doping, in the p and extended s-channels. \cite{note}
Of these, the most
interesting  component is that with extended s-symmetry, for in this
case
$\tilde \Phi_{2 \bf k}$
has the product symmetry $s \otimes d_{x^2 -y^2}= d_{x^2 - y^2}$,
which has even parity and can support singlet pairing.
A careful calculation  gives 
\bea
\tilde \Phi_{2 \bf k}= [cos (k_y)
-\cos (k_x) ]/ (2 \Phi_{1 \bf k})
\eea
Since $J_2/J_1 \sim (x/8)$, 
this 
is a small, but significant perturbation to the  model.
Were it to lead to a genuine pair instability, 
the microscopic description of the state
that forms would involve the coherent presence of
Zhang-Rice singlets of two distinct symmetries.
This is a topic  we shall 
return to in a forthcoming paper.\cite{tobe}

\section{Conclusion}

This paper has examined the effect of interactions around a local
moment.
Conventional
wisdom assumes that a localized  moment scatters electrons in a
symmetry channel of the same local symmetry.
We have shown that electron interactions 
cause 
a local moment to exchange
spin with electrons in scattering channels with different local symmetry.
Close to a Mott transition these effects are
extreme, and the spin-exchange Hamiltonian acquires the symmetry
of a multichannel Kondo problem.
Finally, we have discussed how these new interaction terms
become important in the Kondo lattice, where
the absence of 
a conserved channel index means that second-channel
couplings generate important interactions within the
the low-energy subspace of Kondo singlets.  The possible consequences 
of these new couplings will be analyzed in a subsequent paper.

We should like to thank the ITP, Santa Barbara, where the early
discussions leading to this work took place
under the support of NSF grants DMR-92-23217 and PHYS94-07194.
Discussions with S. Kivelson on related topics during this
period are gratefully acknowledged.
Later work was supported in part by the National Science Foundation
under Grant DMR 96-14999,  the EPSRC, UK and NATO grant CRG. 940040
during a sabbatical stay at  Oxford.
\vskip 0.1 truein
\noindent $^*$ On sabbatical leave from Rutgers University.


\begin{references}
\bibitem{note}In the cuprate metals, there are two in-plane oxygen atoms 
per copper site, so that the doping per oxygen atom is one-half the
overall electron doping, $x= \delta/2$.  From section (III) our
estimate of $J_2$ is thus $J_2 = J_2 \delta/8$.
\end{references}
\end{document}